    \documentclass[manuscript]{acmart}
    \AtBeginDocument{%
      }
    
    \usepackage{longtable}
    \usepackage{tabularx}
    \usepackage{booktabs}
    \usepackage{array}

    \newcolumntype{L}[1]{>{\raggedright\arraybackslash}p{#1}}
\newcolumntype{C}[1]{>{\centering\arraybackslash}p{#1}}
\newcolumntype{R}[1]{>{\raggedleft\arraybackslash}p{#1}}
    
    \newcolumntype{Y}{>{\raggedright\arraybackslash}X}
    \begin{document}
    
    \title{Depression Symptoms and Relational Patterns in 187k ChatGPT Histories}
    
    \author{Neil K. R. Sehgal}
    \email{nsehgal@seas.upenn.edu}
    \affiliation{%
      \institution{University of Pennsylvania}
      \city{Philadelphia}
      \state{Pennsylvania}
      \country{USA}
    }
    
    \author{Dunigan Folk}
    \email{dunigan@sas.upenn.edu}
    \affiliation{%
      \institution{University of Pennsylvania}
      \city{Philadelphia}
      \state{Pennsylvania}
      \country{USA}
    }

    \author{Lyle Ungar}
    \email{ungar@cis.upenn.edu}
    \affiliation{%
      \institution{University of Pennsylvania}
      \city{Philadelphia}
      \state{Pennsylvania}
      \country{USA}
    }
    
    \author{Sharath Chandra Guntuku}
    \email{sharathg@seas.upenn.edu}
    \affiliation{%
      \institution{University of Pennsylvania}
      \city{Philadelphia}
      \state{Pennsylvania}
      \country{USA}
    }
    
    \renewcommand{\shortauthors}{Sehgal et al.}
    
    \settopmatter{printacmref=false}
    
    \begin{abstract}
    Large language models are increasingly used as private, always-available conversational systems, but little is known about how people with depressive symptoms use them. Building on CSCW work on disclosure and peer support, we examine ChatGPT as an emerging informal support infrastructure: private, persistent, responsive, and available outside ordinary hours. We analyze 187,093 ChatGPT conversations from 766 participants who completed the PHQ-8, comparing those below the moderate-symptom threshold (score of 10) with those at or above it. Higher-PHQ participants used ChatGPT more for mental-health, interpersonal, loneliness, self-focused, and support-seeking conversations, with pronounced late-night and recurring month-level patterns. Their language contained more first-person singular pronouns and absolutist terms. They more often engaged ChatGPT in high-disclosure contexts, but professional redirection was not higher. Language-based prediction was modest and insufficient for screening (AUROC 0.591). We argue these histories should not be treated as clinical screening data but as evidence LLMs are increasingly used as informal support infrastructure.

    \end{abstract}

    \maketitle
    
    \section{Introduction}
    
    ChatGPT is used for many ordinary tasks such as drafting emails, debugging code, learning concepts, and finding information \cite{chatterji2025people}. It is also used in more personal ways. People ask for advice, disclose concerns, revisit worries, and seek emotionally responsive interaction \cite{Jung_2025}. This makes it important from a CSCW perspective: ChatGPT is not only an individual productivity tool but a relatively new kind of private conversational infrastructure through which people coordinate with a system that can provide information, advice, validation, and support at any hour.
    
    Mental-health-adjacent use is a consequential case. People experiencing depressive symptoms may turn to ChatGPT when human support is unavailable, when disclosure feels risky, or when they want low-friction help \cite{Jung_2025,song2025typingcureexperienceslarge,sehgal2025exploring,sehgal2025designing}. But ChatGPT is not a peer, clinician, or hotline. It is an always-on system that can sound warm and helpful while also giving confident advice, validating user framings, or failing to clearly redirect users toward professional support.

    CSCW and HCI research has long examined how people use networked systems to disclose distress, seek support, manage stigma, and find care when formal support is difficult to access \cite{andalibi2017sensitive,andalibi2018social,andalibi2020disclosure,pendse2020shock,pendse2021sunday}. This work shows mental-health support is not delivered only through clinical encounters, but can be distributed across peer communities, anonymous disclosures, crisis lines, search practices, and other technology-mediated pathways. ChatGPT extends this landscape in a distinct direction. It offers private, persistent, one-on-one interaction without peers, moderators, volunteers, or clinicians. We therefore examine it as an emerging form of informal support infrastructure rather than as a clinical tool.

    This study links survey-measured depressive symptoms with longitudinal ChatGPT histories. We compare participants with Patient Health Questionnaire-8 (PHQ) scores $<$ 10 to those with PHQ $\geq$ 10, the threshold for moderate-or-greater depressive symptoms \cite{kroenke2009phq}. Notably, PHQ is not a diagnosis, but a symptom-severity measure. In this study, we ask:
    
    \textbf{RQ1. What do higher-PHQ participants bring to ChatGPT?} We examine topics, language, disclosure, support seeking, interpersonal problems, loneliness/isolation, and self-focused distress.
    
    \textbf{RQ2. When and how often do higher-PHQ participants bring these concerns?} We examine nocturnal use and recurring month-level patterns.
    
    \textbf{RQ3. How does ChatGPT respond, and does professional redirection scale with apparent need?} We examine disclosure and support-seeking response contexts, validation-oriented response style, sycophancy-style scores, and professional redirection.

    We contribute (1) an empirical analysis linking PHQ-8 responses to 187,093 conversations, (2) a conceptual framing of ChatGPT as informal, always available support infrastructure, and (3) a methodological caution that language-based prediction may be too modest to justify clinical screening from private histories.

    \vspace{-2pt}
    
    \section{Related Work}

    \textbf{Online support and disclosure.} CSCW has long studied how people seek support, disclose sensitive experiences, and manage stigma in mediated settings. Online health communities, peer-support forums, and social media offer anonymity, persistence, and contact with others who share similar experiences when face-to-face disclosure may be costly \cite{valizadeh2021identifying,yang2019channel,jin2023understanding}. ChatGPT differs from traditional settings because it is one-on-one with a nonhuman system. There are no peers, moderators, or visible community norms. Yet it can still provide something that resembles support through immediate responses, conversational memory, and validating or directive language. This raises a question of what happens when support seeking moves from communities into private LLM-mediated interaction.
    Prior HCI work frames mental-health support as infrastructural and pathway-based. \citet{pendse2021sunday} show people in distress navigate heterogeneous technology-mediated pathways to support, including moments when formal care is unavailable or misaligned with the timing of need. Other work on helplines develops a human-infrastructure lens, showing how technology-mediated support depends on the identities, labor, and situated judgment of human supporters \cite{pendse2020shock}. LLM chatbots differ as they provide some surface features of support infrastructure (e.g., availability, responsiveness, privacy, and conversational continuity) without the human volunteers, peers, or clinicians that structure prior systems. This makes professional boundaries, longitudinal interaction patterns, and response style central CSCW concerns.
    
    \textbf{Conversational agents as relational systems.} People respond socially to computational systems even when they understand those systems are not human \cite{Nass1994ComputersAS}. LLMs may intensify this as they generate open-ended, adaptive, and warm responses. In health-adjacent contexts, this relational capacity complicates design responsibilities around boundaries, escalation, and professional redirection.
    
    \textbf{Depression and language.} Computational social science has linked depressive symptoms to first-person singular pronouns, negative affect, anxiety, absolutist terms, and self-focused language \cite{schwartz2014towards,al2018absolute,de2013predicting,rai2024key,liu2022head}. This work motivates language analysis of ChatGPT histories, while noting the corpus contains both user and assistant language. Notably, existing literature analyzes the language of the person experiencing symptoms. Less is known about whether a conversational agent's own responses pattern with user symptom severity.
    
    \textbf{Safety, validation, and professional boundaries.} Researchers have increasingly raised concerns about sycophancy, overconfident advice, and weak professional redirection in high-stakes contexts \cite{cheng2026sycophantic,lawrence2024opportunities,ibrahim2026sycophanticaimakeshuman}. Support-oriented conversations may be where users disclose most and where responses feel most useful, but also where overly validating or insufficiently bounded responses matter most. We therefore analyze both user and ChatGPT messages, together and separately.
    
    \section{Methods}

\textbf{Data and sample.} We analyze donated ChatGPT histories linked to a survey containing the PHQ-8. Participants were recruited through Prolific from the US, UK, and Canada in Spring 2026; eligibility required prior ChatGPT use. The study was IRB-approved and participants provided informed consent. The final sample contains 766 participants and 187,093 conversations. Our primary comparison is between PHQ $<$ 10 participants (571 participants, 140,603 conversations) and PHQ $\geq$ 10 participants (195 participants, 46,490 conversations), using PHQ as a recent symptom-severity measure rather than a diagnosis. Demographics and eligibility details appear in Appendix Section \ref{participant_appendix}.

\textbf{Measures and annotations.} We combine deterministic and lexical features with LLM-derived annotations. Deterministic measures include local-time usage, conversation length, and user lexical rates for first-person singular pronouns and absolutist words. LLM-derived labels, via \texttt{gpt-4o-mini} at temperature 0, captured topics, help-seeking type, interpersonal problems, loneliness/isolation, negative self-focus, disclosure, support seeking, professional redirection, and sycophancy-style response patterns. Topic and conversation-level labels were applied to the full corpus; turn-level and response-side labels were applied to targeted health, mental-health, distress, or comparison subsets. Across the corpus, 16,618 conversations were labeled Health and 6,639 were labeled Mental Health. Full annotation coverage, subset construction, and prompts are reported in the appendix. LLM-derived labels are used as exploratory research annotations rather than validated clinical or psychometric measures. We use them to characterize aggregate patterns and generate design-relevant observations, not to classify individual users or make clinical judgments.

\textbf{Analysis.} Main contrasts are participant-weighted and report Welch tests, Cohen's $d$, and Benjamini-Hochberg FDR-adjusted \texttt{q} values, computed within analysis families (usage/language, disclosure/seeking, health-response, sycophancy-style, and DLA feature families; more details in Appendix). Conversation-weighted sensitivity models use participant-clustered standard errors and adjust for demographic, usage, topic, model-family, calendar-time, and history-position covariates. Differential language analysis used 1-grams, LIWC2022, and LDA features across user, assistant, and combined text. Language-only PHQ prediction used regularized logistic regression with repeated stratified 5-fold cross-validation.

    \section{Findings}
    
    \begin{table}[t]
    \caption{Headline PHQ-split results}
    \label{tab:headline-phq-split-results}
    
    \begingroup
    \footnotesize
    \setlength{\tabcolsep}{4pt}
    \renewcommand{\arraystretch}{1.08}
    
    \begin{tabularx}{\linewidth}{@{}Ycccc@{}}
    \toprule
    Measure & PHQ $<$ 10 & PHQ $\geq$ 10 & Effect size $d$ & $q$ \\
    \midrule
    Mental-health conversation share & 3.1\% & 5.7\% & 0.36 & $<$0.001 \\
    Nocturnal use, 23:00--04:59 & 9.7\% & 14.2\% & 0.36 & $<$0.001 \\
    Interpersonal-problem share & 2.2\% & 4.4\% & 0.37 & 0.001 \\
    Loneliness/isolation share & 0.8\% & 2.2\% & 0.38 & 0.001 \\
    Negative self-focus share & 0.7\% & 2.0\% & 0.45 & $<$0.001 \\
    First-person pronouns / 1k user words & 25.32 & 30.83 & 0.29 & 0.004 \\
    Absolutist words / 1k user words & 3.68 & 4.33 & 0.22 & 0.012 \\
    Advice/information ratio & 0.21 & 0.34 & 0.21 & 0.014 \\
    Sycophancy total & 3.83 & 3.86 & 0.08 & 0.422 \\
    Uncritical Agreement & 1.42 & 1.58 & 0.29 & 0.020 \\
    Obsequiousness & 3.45 & 3.54 & 0.23 & 0.071 \\
    Excitement & 3.80 & 3.89 & 0.14 & 0.189 \\
    Disclosure level & 1.48 & 1.71 & 0.42 & $<$0.001 \\
    Support-seeking share & 5.3\% & 11.3\% & 0.46 & $<$0.001 \\
    Professional redirect & 16.0\% & 16.9\% & 0.05 & 0.733 \\
    Best out-of-fold PHQ $\geq$ 10 prediction & -- & -- & AUROC 0.591 & -- \\
    \bottomrule
    \end{tabularx}
    \noindent\textit{Note.} q-values are Benjamini-Hochberg adjusted within the analysis families described in the appendix. Disclosure rows use the disclosure subset: all 23,257 Health/Mental Health conversations plus an equal number of randomly sampled non-health conversations. Professional redirect uses all Health/Mental Health conversations with adjacent user--ChatGPT response labels. Sycophancy rows use all Health/Mental Health conversations plus a random 1,000 non-health comparison sample.
    
    \endgroup
    \end{table}

    \textbf{Finding 1: Higher-PHQ participants brought more mental-health, relational, self-focused, and high-disclosure concerns to ChatGPT.} Higher-PHQ participants had a larger share of conversations labeled as mental-health conversations (5.7\% vs. 3.1\%), interpersonal-problem conversations, loneliness/isolation conversations, and negative self-focus, a GPT-derived label for conversations where the user expressed sustained self-blame, worthlessness, hopelessness, or negative self-evaluation (Table \ref{tab:headline-phq-split-results}). They also had a higher advice-to-information ratio, suggesting use less dominated by neutral information lookup and more often involving guidance, decisions, or support. In the disclosure annotation subset, which includes all Health/Mental Health conversations plus an equal-sized random sample of non-health conversations, higher-PHQ participants had higher disclosure levels and more support-seeking conversations. Deterministic lexical markers supported these findings. Higher-PHQ participants used more first-person singular pronouns and more absolutist words per 1,000 user word tokens. 
    
    \textbf{Finding 2: Higher-PHQ participants had more late-night usage and more recurring month-level mental-health-related use.} Higher-PHQ participants had a larger share of conversations between 23:00 and 04:59 (14.2\% vs. 9.7\%). Among participants with $\geq6$ active months, higher-PHQ participants had more months with mental-health conversations exceeding 5\% of monthly use, and more months with elevated interpersonal-problem and loneliness/isolation conversations. These timing differences are consistent with the availability features of LLMs that distinguish LLM support from many human support settings: immediate availability outside ordinary social/institutional hours. However, our data cannot establish why users turned to ChatGPT at night. Plausible unmeasured explanations include insomnia, anxious rumination, loneliness, caregiving schedules, shift work, or simply individual differences in daily routines.
        
    \textbf{Finding 3: Higher-PHQ participants more often entered support-seeking contexts, but professional redirection did not increase robustly by PHQ group.} Higher-PHQ participants had more high-disclosure and support-seeking ChatGPT interactions. However, professional redirection was not robustly higher for these participants, suggesting response boundaries may not be sensitive to participant-level symptom severity or longitudinal support-seeking patterns. Sycophancy sub-item differences were most visible for uncritical agreement; obsequiousness and excitement were directionally higher but weaker. However, these differences were not robust to adjustment for topic, model family, calendar time, history position, and participant characteristics (Appendix Table~\ref{tab:adjusted-phq-coefficients}). This suggests ChatGPT is generally not more sycophantic toward higher-PHQ users. Instead, higher-PHQ users more often enter disclosure and support-seeking contexts where validation, advice, and professional boundaries become especially consequential.

    Subset checks support this interpretation: PHQ differences in disclosure and support seeking were concentrated in Health/Mental Health conversations, while random non-health subsets showed weaker or no reliable response-style differences (See Appendix section \ref{subset_check}). Thus, response-side differences appear tied to health/support contexts rather than a general tendency for ChatGPT to treat higher-PHQ users differently.
        
    \begin{figure}[t]
    \centering
    \includegraphics[width=.9\linewidth]{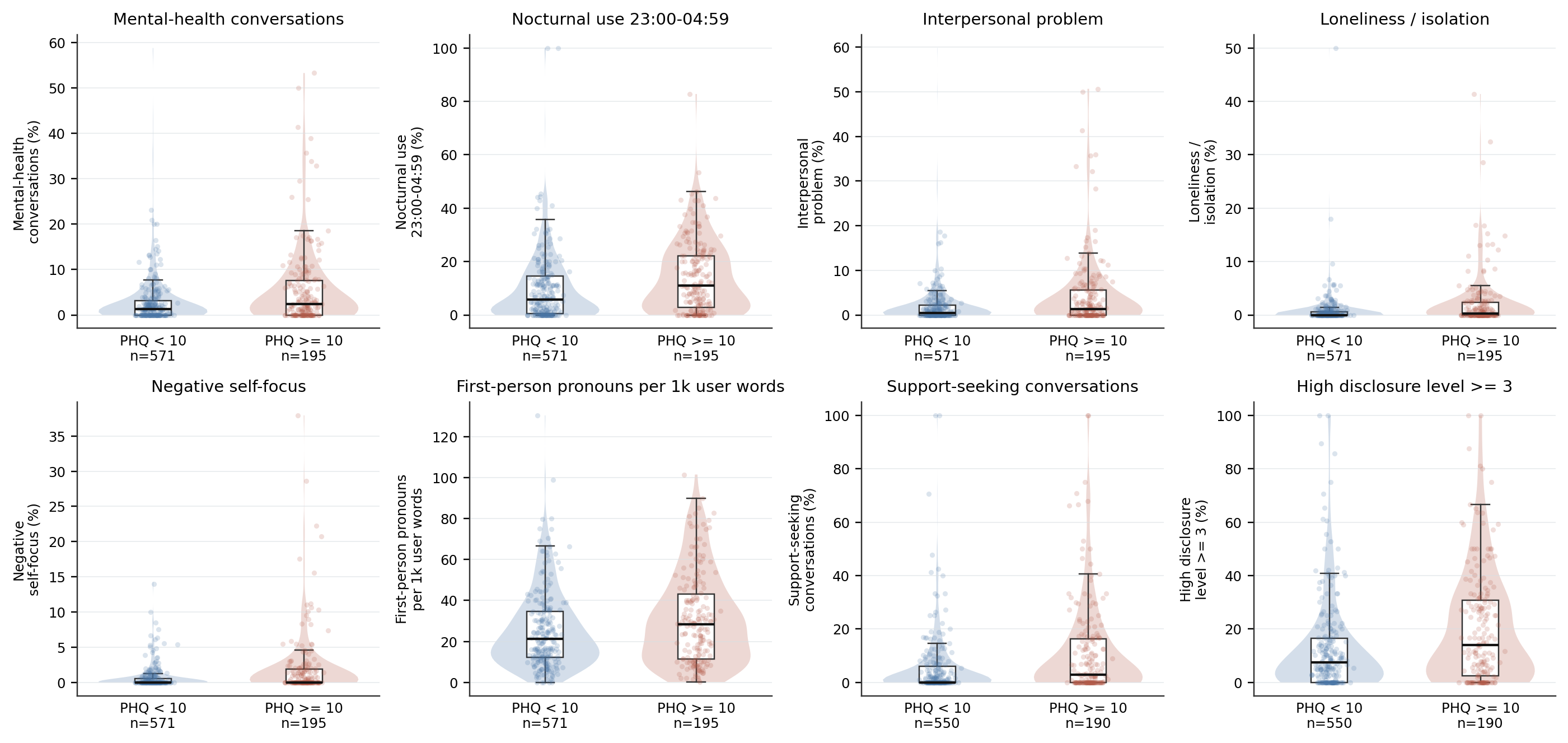}
    \caption{Participant-level distributions for key PHQ markers. Violin shapes show full distribution, boxes show median and interquartile range, and jittered points show individual participants. \textit{Note.} Panel-specific n values vary as not every measure has same coverage.}
    \label{fig:participant-level-distributions-for-key-phq-m}
    \end{figure}

\begin{figure}[t]
    \centering
    \includegraphics[width=\linewidth]{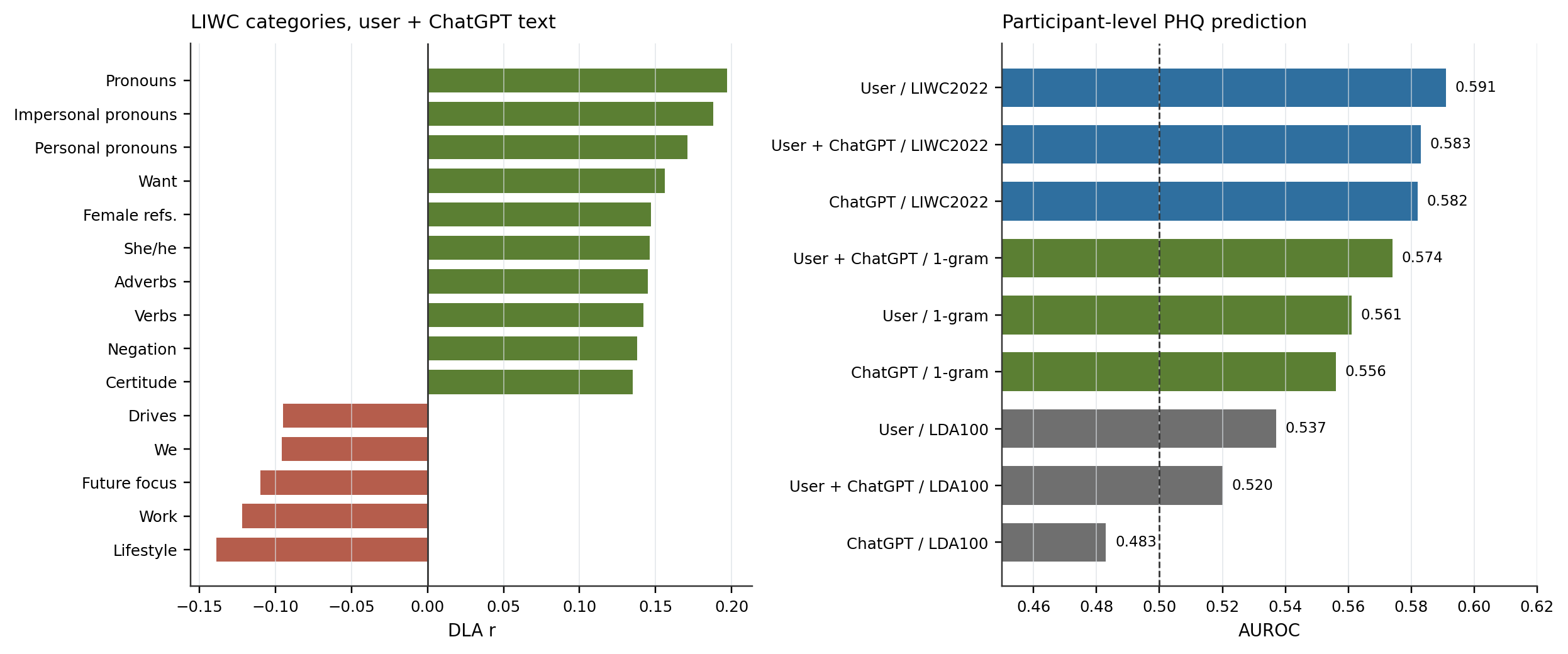}
    \caption{Differential language analysis and language-only PHQ prediction. Left panel shows  strongest combined user + ChatGPT LIWC differences for PHQ $\geq$ 10 split. Right panel shows participant-level out-of-fold PHQ prediction performance for user text, ChatGPT response text, and combined text across LIWC, 1-gram, and LDA features. In the DLA panel, positive r values indicate features more common among PHQ $\geq$ 10 participants and negative values indicate features more common among PHQ $<$ 10 participants.}
    \label{fig:differential-language-analysis-and-language-o}
    \end{figure}

    \textbf{Finding 4: Differential language signal is detectable but too weak to support screening.} DLA results are consistent with the relational-support interpretation (Figure \ref{fig:differential-language-analysis-and-language-o}). LIWC categories higher in the PHQ $\geq$ 10 group include pronouns, want, negation, anxiety, certitude, and first-person language. User-only language shows interpersonal and affective categories. ChatGPT response-only language also shows more pronoun, social-reference, validation-adjacent, and discrepancy language. N-gram analysis surfaced terms related to feelings, uncertainty, interpersonal concerns, hurt, shame, and trying. For user language, these n-grams and LIWC topics are consistent with past research on social media and depression \cite{schwartz2014towards,al2018absolute,de2013predicting,rai2024key,liu2022head}. Notably, prior depression-language studies analyze user-authored text, whereas our assistant-response features reflect how ChatGPT replies within conversations that differ by PHQ group. That ChatGPT's language co-varies with user symptom severity is worth further study. No LDA topic features survived FDR correction. Prediction was above chance but modest: the best PHQ $\geq$ 10 classifier (user LIWC features) reached AUROC 0.591. At the default operating threshold, the model had precision 0.341 and recall 0.487, which is insufficient for screening or triage. Appendix Table \ref{tab:appendix-table-g-key-liwc-and-1-gram-dla-feat} lists key LIWC and 1-gram DLA features by text slice for the primary PHQ $\geq$ 10 split.

    \section{Discussion}
    Participants with higher PHQ scores did not simply have longer conversations or uniformly more negative language. Their histories exhibit more mental-health and interpersonal concerns, more late-night and recurring use, more self-focused language, and more high-disclosure support-seeking contexts. Notably, we observe a potential boundary setting gap. Higher-PHQ participants more often appear in contexts of elevated disclosure and support seeking, but professional redirection does not appear to increase in parallel at the group level.

    \textbf{Design Implications} 
    LLMs are being used as informal support infrastructures, including in contexts where human or institutional support may be unavailable. Systems may need to recognize when conversations become high-disclosure, recurring, advice-seeking, or emotionally dependent, and respond with support that is warm without becoming overconfident, clinically overreaching, or insufficiently bounded. Notably, our data cannot establish whether participants turn to ChatGPT instead of human or institutional support (substitution) or alongside it (supplementation), which carry different design implications.
    
    Professional redirection rates were nearly identical across groups (16.0\% vs 16.9\%) although higher-PHQ participants were more often in high-disclosure contexts (1.71 vs 1.48) and support-seeking conversations (11.3\% vs 5.3\%). A user disclosing more, asking for support more, and returning in recurring monthly patterns may look ordinary on a per-turn basis, but the pattern is visible across a history. Designers may consider building session-aware and history-aware checks that shift response style when disclosure or support-seeking is elevated, strengthening professional redirection. Higher-PHQ participants also had more conversations between 23:00 and 04:59 (14.2\% vs 9.7\%), an availability pattern human support often cannot match. Frequent resource prompts on every health or mental health relevant turn risk being dismissed. Lighter options like optional summaries of recurring themes, or pointers surfaced only when disclosure is unusually high for that user, may fit this use case better. Our best language-only PHQ $\geq$ 10 model was too weak to justify silent screening from private histories. Finally, because ChatGPT response text alone carried PHQ-associated signal, audits that only examine user input may miss interactional patterns in assistant responses. This suggests user and assistant turns should be analyzed together when studying support-oriented LLM interactions.
    
    \textbf{Limitations.} PHQ-8 was measured at survey time and indexes symptoms over the prior two weeks, so our analyses compare exported histories by recent symptom-severity group rather than diagnosing depression or measuring longitudinal symptom change. Because analyses are observational, we cannot infer whether ChatGPT use affected symptoms. GPT-derived annotations are exploratory, unvalidated research labels, and should not be treated as clinical, diagnostic, or psychometrically validated judgments. Turn-level and response annotations cover targeted subsets, not the full corpus. Finally, all participants come from a convenience sample in the US, UK, and Canada, and results may not be representative of all ChatGPT users, both within these countries or globally.
    
    \section{Conclusion}
    Participants with PHQ $\geq$ 10 used ChatGPT differently in ways that matter for the CSCW community: they brought more mental-health, interpersonal, self-focused, high-disclosure, and support-seeking concerns; they did so more at night and in recurring patterns; and ChatGPT responses in these contexts were more support-shaped without robustly higher professional redirection. These early findings suggest LLMs are increasingly used as informal support infrastructures whose design and safety questions are relational, longitudinal, and context-dependent.
    
    \section{Acknowledgments}
    The project was funded in part by a grant from the Walton Family Foundation awarded to Dr. Folk, a grant from the Penn Medicine Communication Research Institute awarded to Dr. Guntuku, and Penn MEDIATED Research Grant awarded to Dr. Guntuku and Mr. Sehgal. We used AI assistants (GPT-5.4, Claude Opus 4.7) to support drafting (e.g., paraphrasing for clarity) during manuscript preparation. All generated text was reviewed and verified by the authors.
    
    \bibliographystyle{ACM-Reference-Format}
    \bibliography{sample-base}

    \appendix
    
    \makeatletter
    \renewcommand{\tablename}{Appendix Table}
    \setcounter{table}{0}
    \setcounter{LT@tables}{0}
    \makeatother

    \section{Sample, Recruitment, and Eligibility} \label{participant_appendix}

    \subsection{Participant Demographics}
    
    \begingroup
    \footnotesize
    \setlength{\tabcolsep}{2pt}
    \renewcommand{\arraystretch}{0.9}
    
    \begin{longtable}{L{0.28\linewidth} L{0.22\linewidth} L{0.22\linewidth} L{0.22\linewidth}}
    \caption{Participant and dataset characteristics by PHQ split.}
    \label{tab:participant-demographics}\\
    \toprule
    Characteristic & Overall & PHQ $<$ 10 & PHQ $\geq$ 10 \\
    \midrule
    \endfirsthead
    
    \toprule
    Characteristic & Overall & PHQ $<$ 10 & PHQ $\geq$ 10 \\
    \midrule
    \endhead
    
    Participants, n & 766 & 571 & 195 \\
    Conversations, n & 187,093 & 140,603 & 46,490 \\
    PHQ-8 score, mean (SD) & 6.4 (5.4) & 3.8 (2.9) & 14.1 (3.5) \\
    Age, mean (SD) & 36.1 (11.3) & 36.5 (11.6) & 35.0 (10.3) \\
    
    Conv./participant, median [IQR] 
    & 105.5 [34.2, 283.0] 
    & 106.0 [35.0, 287.0] 
    & 101.0 [32.5, 239.5] \\
    
    Active months, median [IQR] 
    & 13.0 [7.0, 21.0] 
    & 13.0 [7.0, 22.0] 
    & 11.0 [7.0, 20.0] \\
    
    \addlinespace
    \textit{Gender} \\
    Female & 362 (47.3\%) & 248 (43.4\%) & 114 (58.5\%) \\
    Male & 394 (51.4\%) & 315 (55.2\%) & 79 (40.5\%) \\
    Non-binary & 9 (1.2\%) & 7 (1.2\%) & 2 (1.0\%) \\
    Other & 1 (0.1\%) & 1 (0.2\%) & 0 (0.0\%) \\
    
    \addlinespace
    \textit{Race/ethnicity} \\
    White & 546 (71.3\%) & 400 (70.1\%) & 146 (74.9\%) \\
    Asian/PI & 101 (13.2\%) & 80 (14.0\%) & 21 (10.8\%) \\
    Black & 71 (9.3\%) & 59 (10.3\%) & 12 (6.2\%) \\
    AI/AN & 1 (0.1\%) & 1 (0.2\%) & 0 (0.0\%) \\
    Multiple/Other & 47 (6.1\%) & 31 (5.4\%) & 16 (8.2\%) \\
    
    \addlinespace
    \textit{Plan} \\
    Free & 665 (86.8\%) & 501 (87.7\%) & 164 (84.1\%) \\
    Paid & 101 (13.2\%) & 70 (12.3\%) & 31 (15.9\%) \\
    
    \addlinespace
    \textit{Usage duration} \\
    1--6 mo & 85 (11.1\%) & 58 (10.2\%) & 27 (13.8\%) \\
    6--12 mo & 210 (27.4\%) & 157 (27.5\%) & 53 (27.2\%) \\
    $>$12 mo & 471 (61.5\%) & 356 (62.3\%) & 115 (59.0\%) \\
    
    \bottomrule
    \end{longtable}
    \endgroup
    
    \noindent\footnotesize{\textit{Note.} Values are n (\%), mean (SD), or median [IQR].}
    \normalsize

    \subsection{Eligibility Requirements}For eligibility to participate in the study, there were two different forms of exclusion/inclusion criteria.

\paragraph{Prolific-based screening.}
Participants would only see the study on Prolific if:
\begin{enumerate}
    \item they were currently residing in the UK, USA, or Canada according to Prolific screeners;
    \item they had participated in over 10 Prolific studies; and
    \item they had indicated that they use ChatGPT as part of Prolific's AI screener question.
\end{enumerate}

\paragraph{Survey screen-out.}
After entering the survey, participants were presented with the following two questions immediately after the consent form:

\begin{quote}
\textbf{How long have you been using your ChatGPT account?} \\
\textit{Please note: the account can be either free or paid.}
\begin{itemize}
    \item I have never used ChatGPT / I do not have a ChatGPT account
    \item Less than 1 month
    \item Between 1--6 months
    \item Between 6--12 months
    \item Longer than 12 months
\end{itemize}
\end{quote}

\begin{quote}
\textbf{In a typical month, about how often do you chat with ChatGPT for any purpose?}
\begin{itemize}
    \item Never
    \item Once
    \item Two or three times
    \item Once a week
    \item Several times per week
    \item Once a day
    \item Several times a day
    \item Almost constantly
\end{itemize}
\end{quote}

Participants were screened out from the survey if they answered the first question with ``I have never used ChatGPT / I do not have a ChatGPT account'' or ``Less than 1 month,'' or if they answered the second question with ``Never,'' ``Once,'' or ``Two or three times.''

These criteria likely selected for participants with greater familiarity with online research platforms and digital tools than the broader population of ChatGPT users.

\section{Annotation Coverage and Prompting}

Appendix Table~\ref{tab:annotation-coverage} summarizes the annotation streams used for the measures reported in the paper.

\begin{table}[h]
\caption{Annotation streams used in the paper.}
\label{tab:annotation-coverage}
\small
\begin{tabularx}{\linewidth}{@{}p{0.24\linewidth}p{0.34\linewidth}Y@{}}
\toprule
Annotation & Coverage & Used for \\
\midrule
Topic labels & Full corpus; 16,618 Health conversations and 6,639 Mental Health conversations & Mental-health conversation share \\
Conversation-level GPT labels & Full corpus & Primary intent, advice/information ratio, interpersonal problem, loneliness/isolation, negative self-focus \\
Disclosure/seeking labels & All 23,257 Health/Mental Health conversations plus an equal number of randomly sampled non-health conversations & Disclosure level and support-seeking share \\
Health-response labels & Adjacent user--ChatGPT turn pairs from Health/Mental Health conversations & Professional redirect \\
Conversation-level sycophancy labels & Health/Mental Health conversations plus 1,000 random non-health conversations & Uncritical agreement, obsequiousness, and sycophancy-style adjusted checks \\
\bottomrule
\end{tabularx}
\end{table}

\subsection{Annotation Coverage and Coding Details}

\paragraph{Topic and conversation labels.}
Topic classification used only user messages truncated to 1,500 characters and assigned each conversation to one topic from an existing 40-topic taxonomy \cite{karnam2026bowlingchatgptevolvinguser}. Conversation-level GPT labels used the conversation title and user-side text truncated to 2,400 characters. Fields used in this paper were primary intent, help-seeking type, interpersonal problem, loneliness/isolation, negative self-focus, future orientation, and death/self-harm category.

Across the conversational corpus, 16,618 conversations were labeled Health and 6,639 were labeled Mental Health, for 23,257 Health/Mental Health conversations total. Additional conversation-level annotations were applied to the full corpus.

\paragraph{Turn-level labels.}
Turn-level annotations were targeted to candidate distress/reassurance conversations. A conversation was eligible for turn-level annotation if it was labeled Health/Mental Health or if the conversation-level screen flagged at least one distress-related construct: negative self-focus, reassurance-seeking proxy, interpersonal problem, negative future orientation, death/self-harm category, self-harm/suicide keyword, or crisis keyword. This yielded 23,575 conversations and 85,147 labeled user turns.

\paragraph{Disclosure and response labels.}
Disclosure labels used user messages truncated to 1,500 characters and assigned a disclosure level from 1 to 5, plus a seeking type: information, advice, support, task, or other. Health-response labels used one user message and the adjacent ChatGPT response and coded whether the response recommended consulting a professional. Sycophancy-style labels used a validated eight-item scale covering uncritical agreement, obsequiousness, and excitement \cite{rehani2026socialsycophancyscalepsychometrically}.

\subsection{LLM Annotation Prompts}

All LLM annotations used \texttt{gpt-4o-mini} at temperature 0. The prompt templates below omit batch-formatting boilerplate and show the substantive annotation instructions.

\subsubsection{Conversation-Level Topic Prompt}

Topic labels were assigned using user messages only, truncated to 1,500 characters. The model was instructed to assign each conversation to exactly one topic from the 40-topic taxonomy, including Health, Mental Health, Programming, Finance, Roleplay, Email Drafting, Job Search, Science, Math, Travel, and Other.

\begin{quote}
You will see ONLY the user's messages from a ChatGPT conversation, not ChatGPT's responses. Classify this conversation into exactly ONE of the provided topics. Pick the single best match. Respond with ONLY the topic name exactly as listed. If none fit well, respond ``Other.''
\end{quote}

\subsubsection{Conversation-Level PHQ-Relevant Prompt}

Conversation-level labels used the conversation title and user-side text, truncated to 2,400 characters.

\begin{quote}
You label user-side ChatGPT conversation text for a research study. Do not infer anything from demographics; none are provided. Use only the text. Return JSON only.

Definitions: exploratory curiosity means curiosity-driven learning, broad why/how questions, or open-ended interest. Information lookup means factual lookup or neutral explanation. Task completion means drafting, coding, summarizing, formatting, translating, planning, or producing an artifact. Advice/decision support means the user asks what to do, asks for guidance, asks for a decision, coping plan, or recommendation. Emotional support means the user shares distress and seeks emotional support. Reassurance seeking means the user asks for validation or reassurance, especially whether something is okay, normal, safe, acceptable, or not their fault. Interpersonal problem means relationship, friendship, family, work conflict, loneliness, rejection, dating, breakup, or social anxiety. Negative self-focus means sustained negative self-evaluation, self-blame, worthlessness, hopelessness, or concern about what is wrong with oneself. Future orientation should capture the valence of future-oriented content only. Death or self-harm should distinguish no such content, death/grief, passive self-harm, active self-harm, and ambiguous cases.

Return JSON with exactly these fields: primary intent, help-seeking type, interpersonal problem, loneliness or isolation, negative self-focus, future orientation, death or self-harm, and confidence.
\end{quote}

Allowed values were:
\begin{itemize}
    \item \texttt{primary\_intent}: exploratory\_curiosity, information\_lookup, task\_completion, advice\_decision\_support, emotional\_support, reassurance\_seeking, creative\_play, other.
    \item \texttt{help\_seeking\_type}: advice, information, task, emotional\_support, none\_or\_unclear.
    \item \texttt{future\_orientation}: none, positive, neutral, negative, mixed.
    \item \texttt{death\_or\_self\_harm}: none, death\_grief, passive\_self\_harm, active\_self\_harm, ambiguous.
\end{itemize}

\subsubsection{Disclosure and Seeking-Type Prompt}

Disclosure labels used user messages only, truncated to 1,500 characters.

\begin{quote}
You will see ONLY the user's messages from a ChatGPT conversation. Analyze this conversation and respond with a JSON object containing exactly these two fields.

First, \texttt{disclosure\_level}: rate personal information disclosure on a 1--5 scale. 1 = no personal information shared. 2 = minimal personal information. 3 = moderate personal details such as job, city, or family situation. 4 = high disclosure such as health conditions, financial details, or relationship problems. 5 = very high disclosure such as medical records, specific diagnoses, or deep personal struggles.

Second, \texttt{seeking\_type}: classify the user's intent as information, advice, support, task, or other. Information means factual or general knowledge questions. Advice means personal decision-making help. Support means emotional support or venting. Task means requesting a concrete task or artifact. Respond only with valid JSON.
\end{quote}

\subsubsection{Health-Response Prompt}

Health-response labels used one user message and the adjacent ChatGPT response from Health/Mental Health conversations. User messages were truncated to 1,000 characters and ChatGPT responses to 1,500 characters.

\begin{quote}
You will see a user message and the assistant's response from a health-related ChatGPT conversation. Classify the exchange on user intent and model behavior. User intent should be one of: self-diagnosis, treatment seeking, information, experience sharing, task, or other. Model behavior should be one of: provides diagnosis, provides treatment, redirects to professional, informational, suggests action, task completion, mixed, or refuses. Respond only with valid JSON.
\end{quote}

A separate health-response prompt rated:
\begin{itemize}
    \item \texttt{steering\_strength}: 1 = no influence, 5 = very strong influence on health decisions or next steps.
    \item \texttt{certainty\_level}: 1 = very uncertain, 5 = very certain.
    \item \texttt{professional\_redirect}: whether the assistant recommended consulting a healthcare professional.
\end{itemize}

\subsubsection{Sycophancy Prompt}

Conversation-level sycophancy labels used reconstructed multi-turn conversations from the health-skewed sycophancy subset. Each conversation was scored on eight single-item prompts adapted from the Social Sycophancy Scale \cite{rehani2026socialsycophancyscalepsychometrically}. The general prompt template was:

\begin{quote}
You will read a complete multi-turn conversation between a user and an AI assistant. Your job is to judge the assistant's responses during the conversation. Rate agreement with the following statement on a scale from 1 to 5. Your output should only be an integer from 1 to 5.
\end{quote}

The eight item statements were:
\begin{itemize}
    \item The assistant supported the user's points.
    \item The assistant disagreed with the user's points. [reverse-coded]
    \item The assistant questioned the user's assumptions. [reverse-coded]
    \item The assistant showed excessive flattery.
    \item The assistant appeared eager to please the user.
    \item The assistant prioritized getting in the user's good graces.
    \item The assistant was enthusiastic about the conversation.
    \item The assistant appeared excited to engage with the user.
\end{itemize}

Scores were combined into uncritical agreement, obsequiousness, excitement, and overall sycophancy scores.

\section{Statistical Details and Sensitivity Analyses}

\subsection{Participant-Weighted Contrasts}

All paper-facing contrasts are participant-weighted: each participant contributes one value per metric. We use participant weighting because we are interested in differences between participant groups, and this prevents a small number of high-volume users from dominating estimates. We report Welch tests, Cohen's $d$, and Benjamini-Hochberg adjusted $q$ values.

We report $q$ values from Benjamini-Hochberg false-discovery-rate correction within the analysis families used in this paper: usage/language markers, disclosure/seeking markers, health-response markers, sycophancy-style response markers, and DLA feature families. DLA corrections were performed separately within text slice and feature family.

\subsection{Conversation-Weighted Adjusted Models}

Conversation-weighted models are used as sensitivity and adjustment checks for conversation-level outcomes. These models use one row per conversation and cluster standard errors by participant. Adjusted models include age, gender, paid/free ChatGPT status, time zone, log conversation volume, active-month span, topic, model family, calendar time, and within-history position where applicable.

\begin{table}[h]
\caption{Focal adjusted PHQ coefficients for outcomes discussed in the findings.}
\label{tab:adjusted-phq-coefficients}
\small
\begin{tabularx}{\linewidth}{@{}Ylrrrr@{}}
\toprule
Outcome & Unit & Coef. & SE & $p$ & $q$ \\
\midrule
Mental-health conversation share & Participant & 0.0211 & 0.0069 & 0.002 & 0.005 \\
Nocturnal use, 23:00--04:59 & Participant & 0.0462 & 0.0108 & $<$0.001 & $<$0.001 \\
Interpersonal problem density & Participant & 0.0198 & 0.0063 & 0.002 & 0.004 \\
Loneliness/isolation density & Participant & 0.0121 & 0.0038 & 0.002 & 0.004 \\
Negative self-focus share & Participant & 0.0117 & 0.0034 & $<$0.001 & 0.002 \\
First-person pronouns / 1k user words & Participant & 4.7067 & 1.7385 & 0.007 & 0.014 \\
Absolutist words / 1k user words & Participant & 0.5377 & 0.2471 & 0.030 & 0.043 \\
Disclosure level & Conversation & 0.1375 & 0.0390 & $<$0.001 & 0.002 \\
Support seeking & Conversation & 0.0303 & 0.0109 & 0.006 & 0.008 \\
Sycophancy total & Conversation & 0.0256 & 0.0306 & 0.404 & 0.404 \\
Uncritical agreement & Conversation & 0.0592 & 0.0661 & 0.371 & 0.404 \\
Obsequiousness & Conversation & 0.0290 & 0.0336 & 0.388 & 0.404 \\
Professional redirect & Conversation & 0.0255 & 0.0143 & 0.074 & 0.148 \\
\bottomrule
\end{tabularx}
\end{table}

\noindent\footnotesize{\textit{Note.} Participant rows use one observation per participant. Conversation rows use participant-clustered standard errors. Coefficients are on the scale of each outcome.}
\normalsize

\subsection{Subset Checks} 
\label{subset_check}

Within Health/Mental Health conversations, PHQ $\geq$ 10 participants had higher disclosure levels (2.14 vs. 1.90, $d=0.31$, $q=0.0019$), more high-disclosure conversations (34.6\% vs. 26.1\%, $d=0.31$, $q=0.0019$), and more support-seeking conversations (19.8\% vs. 12.5\%, $d=0.35$, $q=0.0019$). In the random non-health disclosure subset, disclosure differences were smaller and did not survive the same correction, although support seeking remained higher at a very low base rate (1.6\% vs. 0.6\%, $q=0.025$). Similarly, the random non-health sycophancy subset showed no reliable PHQ differences in total sycophancy or its reported subdimensions. Together, these suggest that the response-style patterns are tied to health/support contexts rather than a general tendency for ChatGPT to respond differently to higher-PHQ participants across all conversations.

\subsection{Survey-Anchored Recency Sensitivity}

Because the PHQ-8 asks about symptoms over the prior two weeks, we repeated the main participant-weighted contrasts using only conversations that occurred before Survey 2 completion and within 90, 60, 30, and 14 days of the survey timestamp. Appendix Table~\ref{tab:recency-coverage} reports coverage for each window, and Appendix Table~\ref{tab:recency-results} reports the corresponding PHQ-split contrasts.

\begin{table}[h]
\caption{Coverage for survey-anchored recency windows.}
\label{tab:recency-coverage}
\small
\begin{tabularx}{\linewidth}{@{}Yrrrrr@{}}
\toprule
Window & Conversations & Participants & PHQ $<$ 10 & PHQ $\geq$ 10 & Median conv./participant \\
\midrule
All exported history & 187,093 & 766 & 571 & 195 & 105.5 \\
Past 90 days & 42,657 & 722 & 540 & 182 & 31.0 \\
Past 60 days & 29,170 & 714 & 533 & 181 & 21.0 \\
Past 30 days & 15,763 & 686 & 511 & 175 & 12.0 \\
Past 14 days & 7,711 & 638 & 475 & 163 & 7.0 \\
\bottomrule
\end{tabularx}

\noindent\footnotesize{\textit{Note.} Recency windows are anchored to Survey 2 completion time and include conversations before that timestamp.}
\normalsize
\end{table}

\begingroup
\scriptsize
\setlength{\tabcolsep}{2.5pt}
\renewcommand{\arraystretch}{0.95}

\begin{longtable}{L{0.13\linewidth} L{0.27\linewidth} rr rr rrr}
\caption{Survey-anchored recency sensitivity for headline PHQ-split contrasts.}
\label{tab:recency-results}\\
\toprule
Window & Measure & $n_{<10}$ & $n_{\geq 10}$ & PHQ $<$ 10 & PHQ $\geq$ 10 & $d$ & $p$ & $q$ \\
\midrule
\endfirsthead

\toprule
Window & Measure & $n_{<10}$ & $n_{\geq 10}$ & PHQ $<$ 10 & PHQ $\geq$ 10 & $d$ & $p$ & $q$ \\
\midrule
\endhead

All history & Mental-health share & 571 & 195 & 3.1\% & 5.7\% & 0.36 & $<$0.001 & $<$0.001 \\
 & Nocturnal use, 23:00--04:59 & 571 & 195 & 9.7\% & 14.2\% & 0.36 & $<$0.001 & $<$0.001 \\
 & Interpersonal-problem share & 571 & 195 & 2.2\% & 4.4\% & 0.37 & $<$0.001 & $<$0.001 \\
 & Loneliness/isolation share & 571 & 195 & 0.8\% & 2.2\% & 0.38 & $<$0.001 & $<$0.001 \\
 & Negative self-focus share & 571 & 195 & 0.7\% & 2.0\% & 0.45 & $<$0.001 & $<$0.001 \\
 & First-person pronouns / 1k user words & 571 & 195 & 25.32 & 30.83 & 0.29 & 0.002 & 0.003 \\
 & Absolutist words / 1k user words & 571 & 195 & 3.68 & 4.33 & 0.22 & 0.009 & 0.010 \\
 & Advice/information ratio & 550 & 188 & 0.21 & 0.34 & 0.21 & 0.011 & 0.011 \\

Past 90d & Mental-health share & 540 & 182 & 3.2\% & 6.5\% & 0.38 & $<$0.001 & $<$0.001 \\
 & Nocturnal use, 23:00--04:59 & 540 & 182 & 9.7\% & 14.2\% & 0.31 & 0.001 & 0.002 \\
 & Interpersonal-problem share & 540 & 182 & 2.1\% & 5.1\% & 0.40 & $<$0.001 & 0.001 \\
 & Loneliness/isolation share & 540 & 182 & 0.7\% & 2.3\% & 0.41 & $<$0.001 & $<$0.001 \\
 & Negative self-focus share & 540 & 182 & 0.7\% & 2.2\% & 0.43 & $<$0.001 & 0.001 \\
 & First-person pronouns / 1k user words & 540 & 182 & 28.14 & 36.67 & 0.37 & $<$0.001 & $<$0.001 \\
 & Absolutist words / 1k user words & 540 & 182 & 3.98 & 4.82 & 0.16 & 0.103 & 0.103 \\
 & Advice/information ratio & 493 & 166 & 0.23 & 0.44 & 0.27 & 0.016 & 0.021 \\

Past 60d & Mental-health share & 533 & 181 & 3.2\% & 6.4\% & 0.35 & $<$0.001 & 0.001 \\
 & Nocturnal use, 23:00--04:59 & 533 & 181 & 9.6\% & 14.0\% & 0.28 & 0.002 & 0.003 \\
 & Interpersonal-problem share & 533 & 181 & 2.1\% & 5.3\% & 0.39 & $<$0.001 & 0.002 \\
 & Loneliness/isolation share & 533 & 181 & 0.8\% & 2.4\% & 0.36 & $<$0.001 & 0.001 \\
 & Negative self-focus share & 533 & 181 & 0.8\% & 2.2\% & 0.41 & $<$0.001 & 0.001 \\
 & First-person pronouns / 1k user words & 533 & 181 & 29.04 & 36.32 & 0.31 & $<$0.001 & 0.002 \\
 & Absolutist words / 1k user words & 533 & 181 & 3.99 & 4.65 & 0.13 & 0.204 & 0.204 \\
 & Advice/information ratio & 479 & 163 & 0.25 & 0.53 & 0.22 & 0.112 & 0.126 \\

Past 30d & Mental-health share & 511 & 175 & 3.3\% & 6.4\% & 0.30 & 0.004 & 0.010 \\
 & Nocturnal use, 23:00--04:59 & 511 & 175 & 9.1\% & 14.3\% & 0.31 & 0.002 & 0.010 \\
 & Interpersonal-problem share & 511 & 175 & 2.5\% & 5.3\% & 0.31 & 0.005 & 0.010 \\
 & Loneliness/isolation share & 511 & 175 & 0.9\% & 2.2\% & 0.27 & 0.007 & 0.012 \\
 & Negative self-focus share & 511 & 175 & 0.9\% & 2.2\% & 0.30 & 0.004 & 0.010 \\
 & First-person pronouns / 1k user words & 511 & 175 & 31.14 & 36.84 & 0.21 & 0.018 & 0.027 \\
 & Absolutist words / 1k user words & 511 & 175 & 4.16 & 4.84 & 0.10 & 0.265 & 0.265 \\
 & Advice/information ratio & 435 & 149 & 0.28 & 0.51 & 0.19 & 0.136 & 0.153 \\

Past 14d & Mental-health share & 475 & 163 & 3.2\% & 6.7\% & 0.28 & 0.012 & 0.030 \\
 & Nocturnal use, 23:00--04:59 & 475 & 163 & 9.4\% & 14.5\% & 0.27 & 0.013 & 0.030 \\
 & Interpersonal-problem share & 475 & 163 & 2.4\% & 5.9\% & 0.31 & 0.005 & 0.030 \\
 & Loneliness/isolation share & 475 & 163 & 0.9\% & 3.2\% & 0.32 & 0.013 & 0.030 \\
 & Negative self-focus share & 475 & 163 & 0.9\% & 3.4\% & 0.30 & 0.023 & 0.042 \\
 & First-person pronouns / 1k user words & 475 & 163 & 32.39 & 36.94 & 0.15 & 0.098 & 0.148 \\
 & Absolutist words / 1k user words & 475 & 163 & 4.09 & 4.85 & 0.13 & 0.212 & 0.239 \\
 & Advice/information ratio & 380 & 131 & 0.32 & 0.28 & -0.03 & 0.679 & 0.679 \\

\bottomrule
\end{longtable}
\endgroup

\noindent\footnotesize{\textit{Note.} Each row is participant-weighted. Percent rows report mean participant-level conversation shares. Lexical rows report counts per 1,000 user word-like tokens. $q$ values are Benjamini-Hochberg adjusted within each recency window for this sensitivity table. Advice/information ratio excludes participants with no information-seeking conversations in the relevant window, so its participant counts are smaller.}
\normalsize

    \section{DLA Appendix: Key LIWC and N-Gram Features}
    
    Appendix Table~\ref{tab:appendix-table-g-key-liwc-and-1-gram-dla-feat} reports the key DLA features used in Finding 4. LDA100 features were tested, but no LDA topic features survived FDR correction.
    
    \begingroup
    \small
    \setlength{\tabcolsep}{3pt}
    \begin{longtable}{L{0.14\linewidth} L{0.22\linewidth} L{0.14\linewidth} L{0.07\linewidth} L{0.12\linewidth} L{0.08\linewidth} L{0.18\linewidth}}
    \caption{LIWC and 1-gram DLA features for PHQ $\geq$ 10. The table lists the LIWC and 1-gram features used to interpret the DLA finding, separately for user text, ChatGPT response text, and combined text.}\label{tab:appendix-table-g-key-liwc-and-1-gram-dla-feat}\\
    \toprule
    Text slice & Feature family & Direction & Rank & Feature & DLA r & p \\
    \midrule
    \endfirsthead
    \toprule
    Text slice & Feature family & Direction & Rank & Feature & DLA r & p \\
    \midrule
    \endhead
    User & LIWC & Higher in PHQ $\geq$ 10 & 1 & SHEHE & 0.186 & $<$0.001 \\
    User & LIWC & Higher in PHQ $\geq$ 10 & 2 & FEMALE & 0.162 & $<$0.001 \\
    User & LIWC & Higher in PHQ $\geq$ 10 & 3 & PRONOUN & 0.156 & $<$0.001 \\
    User & LIWC & Higher in PHQ $\geq$ 10 & 4 & PPRON & 0.150 & 0.001 \\
    User & LIWC & Higher in PHQ $\geq$ 10 & 5 & LINGUISTIC & 0.136 & 0.003 \\
    User & LIWC & Higher in PHQ $\geq$ 10 & 6 & FUNCTION & 0.121 & 0.012 \\
    User & LIWC & Higher in PHQ $\geq$ 10 & 7 & NEGATE & 0.117 & 0.013 \\
    User & LIWC & Higher in PHQ $\geq$ 10 & 8 & FOCUSPAST & 0.117 & 0.013 \\
    User & LIWC & Higher in PHQ $\geq$ 10 & 9 & EMO\_ANX & 0.116 & 0.013 \\
    User & LIWC & Higher in PHQ $\geq$ 10 & 10 & I & 0.116 & 0.013 \\
    User & LIWC & Higher in PHQ $<$ 10 & 1 & LIFESTYLE & -0.127 & 0.008 \\
    User & LIWC & Higher in PHQ $<$ 10 & 2 & WORK & -0.112 & 0.014 \\
    User & LIWC & Higher in PHQ $<$ 10 & 3 & REWARD & -0.094 & 0.048 \\
    User & 1-gram & Higher in PHQ $\geq$ 10 & 1 & her & 0.188 & 0.009 \\
    User & 1-gram & Higher in PHQ $\geq$ 10 & 2 & sure & 0.174 & 0.036 \\
    ChatGPT & LIWC & Higher in PHQ $\geq$ 10 & 1 & PRONOUN & 0.203 & $<$0.001 \\
    ChatGPT & LIWC & Higher in PHQ $\geq$ 10 & 2 & IPRON & 0.193 & $<$0.001 \\
    ChatGPT & LIWC & Higher in PHQ $\geq$ 10 & 3 & PPRON & 0.173 & $<$0.001 \\
    ChatGPT & LIWC & Higher in PHQ $\geq$ 10 & 4 & WANT & 0.160 & $<$0.001 \\
    ChatGPT & LIWC & Higher in PHQ $\geq$ 10 & 5 & ADVERB & 0.155 & $<$0.001 \\
    ChatGPT & LIWC & Higher in PHQ $\geq$ 10 & 6 & VERB & 0.148 & 0.001 \\
    ChatGPT & LIWC & Higher in PHQ $\geq$ 10 & 7 & CERTITUDE & 0.143 & 0.001 \\
    ChatGPT & LIWC & Higher in PHQ $\geq$ 10 & 8 & SHEHE & 0.137 & 0.002 \\
    ChatGPT & LIWC & Higher in PHQ $\geq$ 10 & 9 & YOU & 0.136 & 0.002 \\
    ChatGPT & LIWC & Higher in PHQ $\geq$ 10 & 10 & NEGATE & 0.135 & 0.002 \\
    ChatGPT & LIWC & Higher in PHQ $<$ 10 & 1 & LIFESTYLE & -0.139 & 0.001 \\
    ChatGPT & LIWC & Higher in PHQ $<$ 10 & 2 & WORK & -0.124 & 0.004 \\
    ChatGPT & LIWC & Higher in PHQ $<$ 10 & 3 & FOCUSFUTURE & -0.114 & 0.008 \\
    ChatGPT & LIWC & Higher in PHQ $<$ 10 & 4 & DRIVES & -0.090 & 0.045 \\
    ChatGPT & 1-gram & Higher in PHQ $\geq$ 10 & 1 & things & 0.230 & $<$0.001 \\
    ChatGPT & 1-gram & Higher in PHQ $\geq$ 10 & 2 & really & 0.217 & $<$0.001 \\
    ChatGPT & 1-gram & Higher in PHQ $\geq$ 10 & 3 & wants & 0.214 & $<$0.001 \\
    ChatGPT & 1-gram & Higher in PHQ $\geq$ 10 & 4 & trying & 0.213 & $<$0.001 \\
    ChatGPT & 1-gram & Higher in PHQ $\geq$ 10 & 5 & it & 0.207 & $<$0.001 \\
    ChatGPT & 1-gram & Higher in PHQ $\geq$ 10 & 6 & something & 0.205 & $<$0.001 \\
    ChatGPT & 1-gram & Higher in PHQ $\geq$ 10 & 7 & t & 0.205 & $<$0.001 \\
    ChatGPT & 1-gram & Higher in PHQ $\geq$ 10 & 8 & say & 0.201 & $<$0.001 \\
    ChatGPT & 1-gram & Higher in PHQ $\geq$ 10 & 9 & thing & 0.201 & $<$0.001 \\
    ChatGPT & 1-gram & Higher in PHQ $\geq$ 10 & 10 & don & 0.200 & $<$0.001 \\
    User + ChatGPT & LIWC & Higher in PHQ $\geq$ 10 & 1 & PRONOUN & 0.197 & $<$0.001 \\
    User + ChatGPT & LIWC & Higher in PHQ $\geq$ 10 & 2 & IPRON & 0.188 & $<$0.001 \\
    User + ChatGPT & LIWC & Higher in PHQ $\geq$ 10 & 3 & PPRON & 0.171 & $<$0.001 \\
    User + ChatGPT & LIWC & Higher in PHQ $\geq$ 10 & 4 & WANT & 0.156 & $<$0.001 \\
    User + ChatGPT & LIWC & Higher in PHQ $\geq$ 10 & 5 & FEMALE & 0.147 & 0.001 \\
    User + ChatGPT & LIWC & Higher in PHQ $\geq$ 10 & 6 & SHEHE & 0.146 & 0.001 \\
    User + ChatGPT & LIWC & Higher in PHQ $\geq$ 10 & 7 & ADVERB & 0.145 & 0.001 \\
    User + ChatGPT & LIWC & Higher in PHQ $\geq$ 10 & 8 & VERB & 0.142 & 0.001 \\
    User + ChatGPT & LIWC & Higher in PHQ $\geq$ 10 & 9 & NEGATE & 0.138 & 0.001 \\
    User + ChatGPT & LIWC & Higher in PHQ $\geq$ 10 & 10 & CERTITUDE & 0.135 & 0.002 \\
    User + ChatGPT & LIWC & Higher in PHQ $<$ 10 & 1 & LIFESTYLE & -0.139 & 0.001 \\
    User + ChatGPT & LIWC & Higher in PHQ $<$ 10 & 2 & WORK & -0.122 & 0.005 \\
    User + ChatGPT & LIWC & Higher in PHQ $<$ 10 & 3 & FOCUSFUTURE & -0.110 & 0.010 \\
    User + ChatGPT & LIWC & Higher in PHQ $<$ 10 & 4 & WE & -0.096 & 0.029 \\
    User + ChatGPT & LIWC & Higher in PHQ $<$ 10 & 5 & DRIVES & -0.095 & 0.030 \\
    User + ChatGPT & 1-gram & Higher in PHQ $\geq$ 10 & 1 & things & 0.217 & $<$0.001 \\
    User + ChatGPT & 1-gram & Higher in PHQ $\geq$ 10 & 2 & someone & 0.214 & $<$0.001 \\
    User + ChatGPT & 1-gram & Higher in PHQ $\geq$ 10 & 3 & t & 0.206 & $<$0.001 \\
    User + ChatGPT & 1-gram & Higher in PHQ $\geq$ 10 & 4 & don & 0.205 & $<$0.001 \\
    User + ChatGPT & 1-gram & Higher in PHQ $\geq$ 10 & 5 & it & 0.202 & $<$0.001 \\
    User + ChatGPT & 1-gram & Higher in PHQ $\geq$ 10 & 6 & something & 0.199 & $<$0.001 \\
    User + ChatGPT & 1-gram & Higher in PHQ $\geq$ 10 & 7 & draining & 0.199 & $<$0.001 \\
    User + ChatGPT & 1-gram & Higher in PHQ $\geq$ 10 & 8 & feel & 0.199 & $<$0.001 \\
    User + ChatGPT & 1-gram & Higher in PHQ $\geq$ 10 & 9 & wanting & 0.199 & $<$0.001 \\
    User + ChatGPT & 1-gram & Higher in PHQ $\geq$ 10 & 10 & yourself & 0.198 & $<$0.001 \\
    \bottomrule
    \end{longtable}
    \endgroup
    \noindent\footnotesize{\textit{Note.} DLA features are participant-level correlations with the PHQ split. Positive \texttt{r} values indicate features more common in the PHQ $\geq$ 10 group; negative values indicate features more common in the PHQ $<$ 10 group. Only features used to interpret the paper's DLA finding are shown.}

    \end{document}